\documentstyle[prd,aps,preprint,tighten,floats]{revtex}
\begin{document}
\draft
\preprint{IASSNS-HEP-95/80, PUPT-1566}
\date{October 1995}
\title{BPS Saturated Dyonic Black Holes of $N=8$ Supergravity
Vacua\thanks{Based on talks given by M.C. at the SUSY'95 Conference, May
8-12, 1995, Paris, France and the Conference on S-Duality and Mirror
Symmetry, May 22-26, 1995, Trieste, Italy.}}
\author{Mirjam Cveti\v c$^1$
\thanks{On sabbatic leave from the University of Pennsylvania.
 E-mail address: cvetic@sns.ias.edu}
and Donam Youm$^2$
\thanks{On leave from the University of Pennsylvania.
E-mail addresses: youm@pupgg.princeton.edu; youm@sns.ias.edu}}
\address{$^1$ School of Natural Science\\ Institute for Advanced
Study\\ Olden Lane \\ Princeton, NJ 08540 \\ and \\
$^2$ Physics Department\\ Jadwin Hall \\ Princeton University \\
Princeton, NJ 08544}
\maketitle
\begin{abstract}
{We summarize the results for four-dimensional
Bogomol'nyi-Prasad-Sommerfield (BPS) saturated dyonic black hole
solutions arising in the Kaluza-Klein sector and the three-form field
sector of the eleven-dimensional supergravity on a seven-torus.
These black hole solutions break $3\over 4$ of $N=8$ supersymmetry,
fill out the multipletes of $U$-duality group, and the two classes
of solutions are related to each other by a discrete symmetry
transformation in $E_7$.  Using the field redefinitions of the
corresponding effective actions, we present these solutions in terms
of fields describing classes of dyonic black holes carrying charges
of $U(1)$ gauge fields in Neveu-Schwarz-Neveu-Schwarz and/or
Ramond-Ramond sector(s) of the type-IIA superstring on a six-torus.
We also summarize the dependence of their ADM masses on the asymptotic
values of scalar fields, parameterizing the toroidal moduli space
and the string coupling constant.}
\end{abstract}

\section{Introduction}

Recently, there have been rapid developments in understanding of
non-perturbative symmetries in superstring theories.  These
symmetries, called duality symmetries, relate equivalent vacua
in string theories and, especially, $S$-duality \cite{STRWK,HETEM} and
$U$-duality \cite{UDUAL} relate vacua of weakly coupled theories to
those of strongly coupled ones and may shed light on non-perturbative
structure of string vacua.  Duality symmetries are integer valued,
{\it i.e.}, discrete, due to the Dirac charge quantization rules
\cite{DSZ} and space-time and world-sheet instanton effects
\footnote{We shall suppress the coefficients, {\it i.e.}, the set
$Z$ of integers, upon which the duality groups are defined.  So, for
example, $SL(2)$ means the $S$-duality group $SL(2;Z)$, etc.}.

Ten-dimensional superstring theories compactified on a torus ($T^n$)
possess $T$-duality symmetry \cite{TDUAL,SENS}, which relates equivalent
points in toroidal moduli space \cite{NARAIN} of string vacua.
$T$-duality symmetries are exact symmetries of string vacua to all orders
in string perturbation.  Additionally, superstring theories are
conjectured to have strong-weak coupling duality ($S$-duality)
\cite{STRWK,HETEM}, which is a generalization of electric-magnetic
duality for the case of the system with the dilaton and the axion
(pseudo-scalar) fields \cite{HETEM}.  $S$-duality is a non-perturbative
symmetry, since it transforms the dilaton, whose vacuum expectation
value parameterizes the string coupling constant, non-linearly.

It was recently recognized \cite{SENS,UDUAL,WITTEN} that
non-perturbative, BPS saturated states \cite{BPS} play an important role in
gathering evidence for non-perturbative duality symmetries of string
vacua.  In particular, along with the perturbative string excitations,
the non-perturbative BPS saturated states are instrumental in
establishing the equivalence of the two dual string vacua at the level
on the full spectrum of states.  In addition, at the points of moduli
space where non-perturbative BPS states become massless
\cite{HULLCONF,STROM,HTII,CYHETS} they play a crucial role in the
full, low-energy string dynamics.

The study of BPS saturated states of string vacua is thus of importance.
The approach is usually within the effective string action, usually
exact in the leading order of $\alpha'$ expansion, only.  There, one
derives the Bogomol'nyi bound for the ADM mass of a class of
non-perturbative configurations, and in some cases obtains the explicit
solution for the BPS saturated states, which saturate the corresponding
Bogomol'nyi bound, by solving the corresponding Killing spinor equations.
In this classical solutions classical charges are replaced with quantized
charges, which are consistent with the Dirac quantization condition
and the charge constraints of the corresponding string theory.
As a next step, it is important to establish the exactness of such
semi-classical solutions to all orders in $\alpha^{\prime}$ expansion
\cite{CHNUL}.  Ultimately, the proper quantization of the corresponding
states is needed.

Within four-dimensional string vacua, progress has been made in
shedding light on BPS saturated states of $N=4$ string vacua
\footnote{For a review, see Ref. \cite{CYCONF} and references therein}.
In particular, an explicit form of a general class of semi-classical,
BPS saturated, spherically symmetric, static dyonic configurations was
constructed \cite{ALLHET}, and parameterized in terms of fields of
heterotic string compactified on a six-torus.

Recently, it was recognized \cite{TOWN2,WITTEN} that eleven-dimensional
supergravity is the strong coupling limit of the type-IIA superstring.
Therefore, one can infer the strong coupling behavior of the
compactified type-IIA superstring from the compactified
eleven-dimensional supergravity.  Hull and Townsend \cite{UDUAL}
conjectured that the type-II superstring compactified on a torus has a
larger symmetry than ($T$-duality) $\times$ ($S$-duality), {\it i.e.},
$SO(6,6) \times SL(2)$, and the integral version of the $E_7$ symmetry
of the $N=8$ supergravity are realized in the full spectrum of the
type-IIA superstring on a torus.  That is, since the $S$-duality symmetry
of the ten-dimensional type-IIA theory does not commute with the
$T$-duality symmetry of the type-IIA string on a torus,
these two dualities generate a larger symmetry called $U$-duality,
{\it i.e.}, $E_7$, which contains $SO(6,6) \times SL(2)$ as a
maximal subgroup.
Since the 16 Ramond-Ramond (RR) $U(1)$ gauge fields couple to the string
through their field strengths only, the fundamental string states of
the type-IIA string carry only 12 $U(1)$ electric charges of the
Neveu-Schwarz-Neveu-Schwarz (NS-NS) sector, which mix among
themselves under $SO(6,6) \times SL(2)$.  However, the $U$-duality
symmetry requires the existence of additional $16+16$ electric and
magnetic charges in the RR sector which  are carried by solitons
\cite{UDUAL}, and transform as an irreducible spinor representation
under $SO(6,6)$.  In Ref. \cite{UDUAL} prescription was given,
how to generate four-dimensional singly charged solitons with respect
to each of 28 $U(1)$ fields in the RR and the NS-NS sectors by
compactifying p-brane solutions in ten dimensions.

In this contribution, we summarize the results of our previous work
\cite{ELEV} on four-dimensional BPS saturated states of effective,
four-dimensional $N=8$ string vacua.  In this case the results are
only partial and pertain to solutions arising from the following two
sectors: either the Kaluza-Klein sector or the three-form field
sector of the eleven-dimensional supergravity compactified on a
seven-torus.  These results, through the field redefinition, also
provide an explicit construction of classes of dyonic black holes
whose charges arise from the RR sector and/or the NS-NS sector(s) of
type-IIA string on a torus.  By finding solutions of the corresponding
Killing spinor equations, we explicitly construct dyonic BPS saturated
states, carrying charges of 28 $U(1)$ fields in the RR and/or the NS-NS
sector(s).  These solutions fill out the multiplets of the $U$-duality group.
Note, however, that some of the configurations are more general, {\it i.e.},
they preserve only $1\over 4$ of supersymmetry, and cannot be obtained
by simply compactifying p-brane solutions in ten dimensions.

In Section II, we shall summarize the eleven-dimensional supergravity
and discuss its compactification on $T^7$.  In Section III, we give the
BPS saturated dyonic black hole solutions of eleven-dimensional
supergravity on a seven-torus.  In Section IV, field redefinitions
relating the fields of eleven-dimensional supergravity on a seven-torus
to those of the type-IIA supergravity on a six-torus are given, and the
corresponding BPS saturated dyonic black holes carrying $U(1)$ charges
of the NS-NS and/or the RR sector are classified.  Conclusions are given
in Section V.

\section{Lagrangian of Eleven-Dimensional Supergravity on Seven-Torus}

The low energy effective theory of massless states (at generic points
of moduli space) of the type-IIA string in ten dimensions can be
described by eleven-dimensional supergravity on $S^1$ \cite{HUQ}.
Eleven-dimensional supergravity can be thought of as the strong coupling
limit of the type-IIA superstring in ten dimensions \cite{WITTEN}.
Therefore, the strong coupling limit behavior of the lower
dimensional type-IIA theory can be deduced from eleven-dimensional
supergravity compactified down to lower dimensions.

Eleven-dimensional supergravity \cite{ELE} contains the Elfbein
$E^{(11)\,A}_M$, gravitino $\psi^{(11)}_M$, and the three-form field
$A^{(11)}_{MNP}$ as its field content.  The bosonic Lagrangian
density is given by
\begin{equation}
{\cal L} = -{1\over 4}E^{(11)}[{\cal R}^{(11)} + {1\over {12}}
F^{(11)}_{MNPQ}F^{(11)\,MNPQ} - {8\over {12^4}}
\varepsilon^{M_1 \cdots M_{11}}F^{(11)}_{M_1 \cdots M_4}
F^{(11)}_{M_5 \cdots M_8}A^{(11)}_{M_9 M_{10} M_{11}}],
\label{action11d}
\end{equation}
where $F^{(11)}_{MNPQ} \equiv 4\partial_{[M}A^{(11)}_{NPQ]}$ is the field
strength associated with the three-form field $A^{(11)}_{MNP}$.
In the bosonic background, the gravitino $\psi^{(11)}_M$ transforms
under supersymmetry as
\begin{equation}
\delta \psi^{(11)}_M = D_M\, \varepsilon +{i\over 144} (\Gamma^{NPQR}_
{\ \ \ \ \ M} - 8\Gamma^{PQR}\delta^N_M)F^{(11)}_{NPQR}\,\varepsilon ,
\label{sstran11d}
\end{equation}
where $D_M\,\varepsilon = (\partial_M + {1\over 4}\Omega_{MAB}
\Gamma^{AB})\,\varepsilon$ is the gravitational covariant derivative
on the spinor $\varepsilon$, and $\Omega_{ABC} \equiv -\tilde{\Omega}_
{AB,C} + \tilde{\Omega}_{BC,A} - \tilde{\Omega}_{CA,B}$
($\tilde{\Omega}_{AB,C} \equiv E^{(11)\,M}_{[A}E^{(11)\,N}_{B]}
\partial_N E^{(11)}_{MC}$) is the spin connection defined in terms of
the Elfbein.

The four-dimensional effective action can be obtained by
compactifying the extra six spatial coordinates on a seven-torus
$T^7$ by using the following Ansatz for the Elfbein:
\begin{equation}
E^{(11)\,A}_M = \left ( \matrix{e^{-{\varphi \over 2}} e^{\alpha}_{\mu} &
B^i_{\mu} e^a_i \cr 0 & e^a_i} \right ),
\label{elfbein}
\end{equation}
where $\varphi \equiv {\rm ln}\,{\rm det}\,e^a_i$ and
$B^i_{\mu}$ ($i = 1,...,7$) are Kaluza-Klein Abelian gauge fields.
Eleven-dimensional supergravity on $T^7$ has $E_7$ on-shell symmetry
and 70 scalar fields take values in the moduli space
$E_7/[SU(8)/Z_2]$ \cite{CRE}.  However, here we shall write down
$SL(7)$ symmetric truncation of the Lagrangian for the purpose
of obtaining black hole solutions carrying charges of Kaluza-Klein
$U(1)$ or three-form $U(1)$ fields.  These $U(1)$ fields will later
on be related to various $U(1)$ gauge fields of the RR and the NS-NS
sectors of the type-IIA superstring in order to obtain the
corresponding dyonic solitons carrying $U(1)$ charges of the RR and/or
the NS-NS sectors.  The Lagrangian is given by
\begin{equation}
{\cal L} = -{1\over 4}e[{\cal R} - {1\over 2}\partial_{\mu} \varphi
\partial^{\mu} \varphi +{1\over 4}\partial_{\mu} g_{ij} \partial^{\mu}
g^{ij} - {1\over 4} e^{\varphi}g_{ij}G^i_{\mu\nu}G^{j\,\mu\nu}
+{1\over 2}e^{\varphi}g^{ik}g^{jl}F^4_{\mu\nu\,ij}
F^{4\,\mu\nu}_{\ \ \ \ kl} + \cdots ],
\label{action4d}
\end{equation}
where $G^i_{\mu\nu} \equiv \partial_{\mu} B^i_{\nu} - \partial_{\nu}
B^i_{\mu}$, $F^4_{\mu\nu\,ij}\equiv F^{\prime}_{\mu\nu\,ij} +
G^k_{\mu\nu}A_{ijk}$, and the dots ($\cdots$) denote the terms involving
the pseudo-scalars $A_{ijk}$ and the two-form fields $A_{\mu\nu\,i}$.
Here, $F^{\prime}_{\mu\nu\, ij}$ is the field strength of
$A^{\prime}_{\mu\,ij} \equiv A_{\mu\,ij}-B^k_{\mu}A_{kij}$, which
are canonical four-dimensional $U(1)$ fields that are defined to be
scalars under the internal coordinate transformations $x^i \to
x^{\prime\, i} = x^i + \xi^i$ and transform as a $U(1)$ field under
$\delta A^{(11)}_{MNP} = \partial_M \zeta_{NP} + \partial_N \zeta_{PM} +
\partial_P \zeta_{MN}$.  The four-dimensional effective action
(\ref{action4d}) has symmetry under the following $SL(7)$ target space
transformations:
\begin{equation}
g_{ij} \to U_{ik} g_{kl} U_{jl},\ \ \ \
G^i_{\ \mu\nu} \to (U^{-1})_{ik} G^k_{\ \mu\nu} , \ \ \ \
F^4_{\mu\nu\,ij} \to (U^{-1})_{ik} (U^{-1})_{jl} F^4_{\mu\nu\,kl},
\label{GL7}
\end{equation}
and the dilaton $\varphi$ and the four-dimensional metric $g_{\mu\nu}$
remain intact, where $U \in SL(7)$.

\section{Supersymmetric Dyonic Solutions of Eleven-Dimensional
Supergravity}

Given the $SL(7)$ symmetric effective four-dimensional theory, we would
like to solve static, spherically symmetric solutions carrying either
Kaluza-Klein or three-form $U(1)$ gauge field.  The spherically symmetric
Ansatz for the four-dimensional space-time metric is chosen to be
\begin{equation}
g_{\mu\nu}dx^{\mu}dx^{\nu} = \lambda(r)dt^2  -
\lambda^{-1}(r)dr^2 - R(r)(d\theta^2 + {\rm sin}^2 \theta
d\phi^2)
\label{4dmet}
\end{equation}
and the scalar fields depend on the radial coordinate $r$ only.
By solving the Maxwell's equations with the above spherically
symmetric Ans\" atze, one obtains the following non-zero
components of $U(1)$ field strengths:
\begin{eqnarray}
G^i_{tr} &=& {{g^{ij}\tilde{Q}_j} \over {Qe^{\varphi}}},\ \ \ \ \
G^i_{\theta\phi} = P^i {\rm sin}\theta ; \cr
F_{tr\,ij} &=& {{g_{ik}g_{jl}\tilde{Q}^{kl}}\over {Re^{\varphi}}},
\ \ \ \ \ F_{\theta\phi\,ij} = P_{ij}{\rm sin}\theta ,
\label{emfield}
\end{eqnarray}
where the physical electric charges are given by $Q^i =
e^{-\varphi_{\infty}}g^{ij}_{\infty}\tilde{Q}_j$ and $Q_{ij} =
e^{-\varphi_{\infty}}g_{il\,\infty}g_{jl\,\infty}\tilde{Q}^{kl}$.

\subsection{BPS States in the Kaluza-Klein Sector}

The first class of solutions is black holes carrying charges of
Kaluza-Klein $U(1)$ gauge fields.  With the other fields turned off except
the dilaton, the internal metric and Kaluza-Klein $U(1)$ fields, the
four-dimensional effective action reduces to that of eleven-dimensional
Kaluza-Klein theory compactified on a seven-torus.  It was shown
\cite{SUPER} that with a diagonal internal metric, the most general
spherically symmetric configuration corresponds to $U(1)_M \times
U(1)_E$ charged dyonic black hole.  By implementing this solution
with $SO(7)/SO(5)$ transformations \cite{STAT}, one obtains the most
general supersymmetric spherically symmetric Kaluza-Klein black
holes with one constraint \cite{SUPER} $\vec{\cal{P}}\cdot
\vec{\cal{Q}} = 0$, where $\vec{\cal{P}} \equiv (P_1,...,P_7)$
and $\vec{\cal{Q}} \equiv (Q_1,...,Q_7)$ are magnetic and
electric charge vectors of Kaluza-Klein $U(1)$ gauge fields.

Explicit supersymmetric $U(1)_M \times U(1)_E$ solutions \cite{SUPER}
of eleven-dimensional Kaluza-Klein theory with the $j$-$th$ gauge field
magnetic and the $k$-$th$ gauge field electric are given by
\begin{eqnarray}
\lambda &=& {{r - |{\bf P}_{j\,\infty}| - |{\bf Q}_{ k\,\infty}|} \over
{(r - |{\bf P}_{ j\,\infty}|)^{1\over 2}(r - |{\bf Q}_{ k\,\infty}|)^
{1\over 2}}},\ \ \ R\lambda=r^2,
\ \ e^{{2 }({\varphi} - {\varphi}_{\infty})} =
{{r - |{\bf P}_{j\,\infty}|} \over {r - |{\bf Q}_{k\,\infty}|}},
\nonumber \\
{{g_{KK\,ii}}\over {g_{KK\,ii\,\infty}}} &=&
1\ \ (i \neq j,k),
\ \ {{g_{KK\,jj}}\over{g_{KK\,jj\,\infty}}}= {{r - |{\bf P}_{j\,\infty}| -
|{\bf Q}_{k\,\infty}|} \over {r - |{\bf Q}_{k\,\infty}|}},
\ \ {g_{KK\,kk}\over g_{KK\,kk\,\infty}} = {{r - |{\bf P}_{j\,\infty}|}
\over {r - |{\bf P}_{j\,\infty}| - |{\bf Q}_{k\,\infty}|}},
\label{Kaluza-Kleinsol}
\end{eqnarray}
where ${\bf P}_{j\,\infty} \equiv e^{{1\over 2}{\varphi}_{\infty}}
g^{1\over 2}_{KK\,jj\,\infty}P_j$ and ${\bf Q}_{k\,\infty} \equiv
e^{{1\over 2}{\varphi}_{\infty}}g^{1\over 2}_{KK\,kk\,\infty}Q_k$
are the ``screened'' magnetic and electric charges.  Here, the
subscript $\infty$ denotes the asymptotic ($r\to \infty$) value
of the corresponding field and the ADM mass of the configuration
is given by $M=|{\bf P}_{j\,\infty}| + |{\bf Q}_{k\,\infty}|$.

\subsection{BPS States Carrying the Three-Form $U(1)$ Charges}

The second class of solutions corresponds to black holes carrying
charges of the three-form $U(1)$ gauge fields, {\it i.e.}, the $U(1)$
gauge fields obtained from the dimensional reduction of the three-form
field $A^{(11)}_{MNP}$.  With all the other fields, except the
dilaton, the internal metric and the three-form $U(1)$ gauge fields,
turned off, the four-dimensional Lagrangian density reduces to
\begin{equation}
{\cal L} = -{1\over 4}e[{\cal R} - {1\over 2}\partial_{\mu} \varphi
\partial^{\mu} \varphi + {1\over 4} \partial_{\mu} g_{ij} \partial^{\mu}
g^{ij} + {1\over 2}e^{\varphi} g^{ik}g^{jl}F_{\mu\nu\,ij}
F^{\mu\nu}_{\ \ \,kl}],
\label{action3}
\end{equation}
where $F_{\mu\nu\,ij} \equiv \partial_{\mu}A_{\nu\,ij} -
\partial_{\nu}A_{\mu\,ij}$.  This Lagrangian has a manifest
invariance under $SO(7) \subset SL(7)$ with $g_{ij}$ and
$A_{\mu\nu\,ij}$ transforming as $\bf 27$ symmetric and
$\bf 21$ antisymmetric representations of $SO(7)$, respectively.

Just as in the case of Kaluza-Klein BH's, one can obtain constraints on
charges by solving the Killing spinor equations.  These are given by
\begin{equation}
\sum {\bf P}_{ab} {\bf P}_{cd} \gamma^{abcd} = 0 =
\sum {\bf Q}_{ab} {\bf Q}_{cd} \gamma^{abcd},\ \ \ \ \
\sum_{i \neq j} {\bf P}_{ij} {\bf Q}^{ij} = 0 =
\sum {\bf P}_{ab} {\bf Q}_{cd} \gamma^{abcd},
\label{chcon}
\end{equation}
where ${\bf P}_{ab} \equiv e^{\varphi \over 2}e^i_a e^j_b P_{ij}$,
${\bf Q}_{ab} \equiv e^{-{\varphi \over 2}}e^a_i e^b_j Q^{ij}$,
and $\gamma^a$'s are gamma matrices satisfying $SO(3,1)$ Clifford
algebra.  Analyzing the asymptotic form of the above constraints
and applying subsets of global $SL(7)$ transformations, one can see
that static, spherically symmetric solutions with the most general
three-form $U(1)$ charge configurations can be obtained by applying
$SO(7)/SO(3)$ transformation on the configurations with non-zero
charges given by $P_{ij} \neq 0$ and $Q^{ik} \neq 0$ ($j \neq k$),
where $i$ is a fixed index.  Therefore, the static, spherically
symmetric configurations carrying three-form $U(1)$ charges have
$(3n-6)+2=17$ independent charge degrees of freedom.

Once again, this generating solution with non-zero charges
$P_{ij}$ and $Q^{ik}$ is configuration with a diagonal internal
metric, and is given by
\begin{eqnarray}
\lambda &=& {{r - 2|{\bf P}_{ij\,\infty}| - 2|{\bf Q}_{ik\,\infty}|}
\over {(r - 2|{\bf P}_{ij\,\infty}|)^{1\over 2}(r - 2|{\bf Q}_
{ik\,\infty}|)^{1\over 2}}} , \ \ R\lambda=r^2,
\ \ e^{3(\varphi - \varphi_{\infty})} = \left ( {{r - 2|{\bf P}_
{ij\,\infty}|}\over {r - 2|{\bf Q}_{ik\,\infty}|}}\right)^{1\over 2} ,
\nonumber \\
g_{H\,ii}/g_{H\,ii\,\infty} &=& \left ( {{r - 2|{\bf Q}_{ik\,\infty}|}
\over {r - 2|{\bf P}_{ij\,\infty}|}}\right)^{2\over 3},\ \
g_{H\,jj}/g_{H\,jj\,\infty} = {{(r - 2|{\bf P}_{ij\,\infty}|)^{1\over 3}
(r-2|{\bf Q}_{ik\,\infty}|)^{2\over 3}} \over {r - 2|
{\bf P}_{ij\,\infty}| - 2|{\bf Q}_{ik\,\infty}|}} ,
\nonumber \\
g_{H\,kk}/g_{H\,kk\,\infty} &=&
{{r - 2|{\bf P}_{ij\,\infty}| - 2|{\bf Q}_{ik\,\infty}|} \over
{(r - 2|{\bf P}_{ij\,\infty}|)^{2\over 3} (r-2|{\bf Q}_{ik\,\infty}|)^
{1\over 3}}} ,\ \
g_{H\,\ell \ell}/ g_{H\,\ell \ell \,\infty} = \left ( {{r - 2
|{\bf P}_{ij\,\infty}|}\over {r - 2|{\bf Q}_{ik\,\infty}|}}
\right)^{1\over 3} \  (\ell \neq i,j,k),
\label{antisol}
\end{eqnarray}
where ${\bf P}_{ij\,\infty} \equiv e^{\varphi_\infty/2}
g^{-{1\over 2}}_{H\,ii\,\infty}g^{-{1\over 2}}_{H\,jj\,\infty}P_{ij}$
and ${\bf Q}_{ik\,\infty} \equiv e^{\varphi_\infty/2}
g^{-{1\over 2}}_{H\,ii\,\infty} g^{-{1\over 2}}_{H\,kk\,\infty}Q_{ik}$.

\subsection{Relation between Kaluza-Klein and Three-Form Black Hole
Solutions}

The above two classes of solutions have the same global space-time
structure.  That is, for non-zero magnetic and electric charges,
the temperature $T_H \propto 1/\sqrt{\bf PQ}$ is finite, the
singularity is null and the entropy is zero.  If either of $P$ or $Q$
is zero, the singularity becomes naked and the temperature diverges.
This can be traced back to the fact that eleven-dimensional supergravity
compactified on a seven-torus has $E_7$ global symmetry \cite{CRE},
which puts all the 28 $U(1)$ gauge fields and 28 duals on the same putting,
and leaves the four-dimensional space-time metric intact.  The discrete
subset of $E_7$ which relates the above two classes of solutions
is given by
\begin{eqnarray}
P_{ij}&=&{P_j \over 2},\ \ \ \ \ \  Q^{ik}={Q^k \over 2}; \ \ \ \ \ \
{{\check{g}_{H\,jj}}\over {\check{g}_{H\,kk}}} =
{\check{g}_{KK\,kk} \over \check{g}_{KK\,jj}},
\nonumber \\
\check{g}_{H\,jj}\check{g}_{H\,kk}=(\check{g}_{KK\,kk}
&\check{g}_{KK\,jj}&)^{-1/3},\ \check{g}_{H\,ii}= (\check{g}_{KK\,jj}
\check{g}_{KK\,kk})^{-2/3},\  \prod_{\ell\neq (ijk)}
\check{g}_{H\,\ell\ell}=(\check{g}_{KK\,jj}\check{g}_{KK\,kk})^{4/3},
\label{trans}
\end{eqnarray}
where $\check{g}_{ij} \equiv g_{ij}/g_{ij\,\infty}$.

By explicitly solving the Killing spinor equations we have shown
that the above two classes of solutions admit Killing spinors,
{\it i.e.}, preserve some of supersymmetries.  The spinor
$\epsilon$ is constrained by one constraint for each type
(electric or magnetic) of non-zero charges.  So, purely electric
[or magnetic] solutions preserve $1 \over 2$ of the original
supersymmetry, while dyonic solutions preserve only $1 \over 4$
of the original supersymmetry.

Note, however, that a set of configurations obtained in the manner
discussed above constitutes only a subset of the most general BPS
saturated configuration.  One can obtain a more general class of
solutions carrying charges of Kaluza-Klein $U(1)$ {\it and} the
three-form $U(1)$ gauge fields, which will turn out to be the
``generating'' solution that yields the most general supersymmetric
black hole solution of eleven-dimensional supergravity on a
seven-torus after imposing a subset of the $E_7$ transformations
\footnote{The corresponding generating solution in the heterotic
string on a six-torus has been obtained \cite{ALLHET}: it carries two
electric [and two magnetic] charges of the Kaluza-Klein $U(1)$ field
and the two-form $U(1)$ field having common indices.  Each type
(electric or magnetic) of charge break $1 \over 2$ of the original
supersymmetry.  Note that such dyonic solutions cannot be obtained
from the singly charged solutions through the $S$-duality
transformations.}.
Such a generating solution would in turn allow us to study the full
spectrum of the U-duality symmetry of the $N=8$ superstring vacua.

\section{BPS states of Type-II String Compactified on Six-Torus}

By using the fact that the type-IIA superstring on a six-torus is
dual to eleven-dimensional supergravity on a seven-torus, one can transform
the above two classes of solutions in eleven-dimensional supergravity on
a seven-torus into those in the type-IIA superstring on a six-torus.
The zero slope limit of the ten-dimensional type-IIA superstring can be
described by the eleven-dimensional supergravity compactified on a
circle $S^1$.  The dimensional reduction is accomplished by the
following choice of the Kaluza-Klein Ansatz for the Elfbein
$E^{(11)\,A}_M$:
\begin{equation}
E^{(11)\, A}_M = \left ( \matrix{e^{-{\Phi \over 3}}
e^{(10)\, \breve{\alpha}}_{\breve{\mu}} & e^{{2\over 3}\Phi}
B_{\breve{\mu}} \cr 0 & e^{{2\over 3}\Phi}} \right ) ,
\label{Kaluza-Kleintoten}
\end{equation}
where $\Phi$ corresponds to the ten-dimensional dilaton field in the
NS-NS sector, $e^{(10)\,\breve{\alpha}}_{\breve{\mu}}$ is the Zehnbein in
the NS-NS sector, and $B_{\breve{\mu}}$ corresponds to a one-form in
the RR sector of superstring.  Here, the breve denotes the ten-dimensional
space-time vector index.  And the three-form $A^{(11)}_{MNP}$ is
decomposed into $A^{(11)}_{MNP} = (A_{\breve{\mu}\breve{\nu}\breve{\rho}},
A_{\breve{\mu}\breve{\nu}11} \equiv A_{\breve{\mu}\breve{\nu}})$,
where $A_{\breve{\mu}\breve{\nu}\breve{\rho}}$ is identified as
the three-form in the RR sector and $A_{\breve{\mu}\breve{\nu}}$ is the
antisymmetric tensor in the NS-NS sector.

Then, the eleven-dimensional bosonic action reduces to the following
effective action for the massless bosonic fields in the type-IIA
superstring:
\begin{equation}
\cal{L} = \cal{L}_{NS} + \cal{L}_R ,
\label{sg10d}
\end{equation}
with
\begin{eqnarray}
{\cal L}_{NS} &=& -{1\over 4}e^{(10)}e^{-2\Phi}[{\cal R} +
4\partial_{\breve{\mu}}\Phi \partial^{\breve{\mu}}\Phi -
{1\over 3}F_{\breve{\mu}\breve{\nu}\breve{\rho}}
F^{\breve{\mu}\breve{\nu}\breve{\rho}}],
\nonumber \\
{\cal L}_R &=& -{1 \over 4}e^{(10)}[{1\over 4}G_{\breve{\mu}\breve{\nu}}
G^{\breve{\mu}\breve{\nu}} +{1\over {12}}F^{\prime}_{\breve{\mu}\breve{\nu}
\breve{\rho}\breve{\sigma}} F^{\prime\, \breve{\mu}\breve{\nu}\breve{\rho}
\breve{\sigma}} - {6 \over {(12)^3}}\varepsilon^{\breve{\mu}_1 \cdots
\breve{\mu}_{10}} F_{\breve{\mu}_1 \cdots \breve{\mu}_4}
F_{\breve{\mu}_5 \cdots \breve{\mu}_8} A_{\breve{\mu}_9 \breve{\mu}_{10}}],
\label{nsr}
\end{eqnarray}
where $F_{\breve{\mu}\breve{\nu}\breve{\rho}} \equiv
3\partial_{[\breve{\mu}}A_{\breve{\nu}\breve{\rho}]}$,
$G_{\breve{\mu}\breve{\nu}} \equiv 2\partial_{[\breve{\mu}}
B_{\breve{\nu}]}$, $F^{\prime}_{\breve{\mu}\breve{\nu}
\breve{\rho}\breve{\sigma}} \equiv 4\partial_{[\breve{\mu}}
A_{\breve{\nu}\breve{\rho}\breve{\sigma}]}
-4F_{[\breve{\mu}\breve{\nu}\breve{\rho}}B_{\breve{\sigma}]}$,
and $\varepsilon^{\breve{\mu}_1\cdots \breve{\mu}_{10}} \equiv
\varepsilon^{\breve{\mu}_1\cdots \breve{\mu}_{10}11}$.  The
eleven-dimensional gravitino $\psi^{(11)}_M$ is decomposed into
the ten-dimensional gravitino $\psi_{\breve{\mu}}$ and a fermion
$\psi_{11}$ as $\psi^{(11)}_M = (\psi_{\breve{\mu}}, \psi_{11})$.
These spinors can be split into two Majorana-Weyl spinors of
left- and right-helicities, thereby describing $N=2$ supergravity.

To compactify the above ten-dimensional effective action on a six-torus down
to four dimensions, one chooses the following Kaluza-Klein Ansatz for
the Zehnbein:
\begin{equation}
e^{(10)\, \breve{\alpha}}_{\breve{\mu}} =
\left ( \matrix{e^{\alpha}_{\mu} & \bar{B}^m_{\mu}\bar{e}^a_m \cr
0 & \bar{e}^a_m} \right ),
\label{zehnbein}
\end{equation}
where $\bar{B}^m_{\mu}$ ($m=1,...,6$) are Abelian Kaluza-Klein gauge fields
(in the NS-NS sector), $e^{\alpha}_{\mu}$ is the string frame Vierbein and
$\bar{e}^a_m$ is the Sechsbein.  Setting all the scalars except those
associated with the Sechsbein $\bar{e}^a_m$ and the ten-dimensional
dilaton $\Phi$ to zero, one has the following string-frame
four-dimensional Lagrangian for the type-IIA superstring on a six-torus:
\begin{eqnarray}
{\cal L}_{II} = &-&{1\over 4}e[e^{-2\phi}({\cal R} + 4\partial_{\mu}
\phi \partial^{\mu}\phi + {1\over 4}\partial_{\mu}\bar{g}_{mn}
\partial^{\mu}\bar{g}^{mn} - {1\over 4}\bar{g}_{mn}\bar{G}^m_{\mu\nu}
\bar{G}^{n\, \mu\nu} -\bar{g}^{mn}{\bar F}_{\mu\nu\, m}
{\bar F}^{\mu\nu}_{\ \ n})
\nonumber \\
&+&{1\over 4}e^{{\bar \sigma}}\bar{G}_{\mu\nu}\bar{G}^{\mu\nu} +
{1\over 2}e^{{\bar \sigma}} \bar{g}^{mn}\bar{g}^{pq}
\bar{F}_{\mu\nu\, mp}\bar{F}^{\mu\nu}_{\ \ nq}] ,
\label{type2}
\end{eqnarray}
where $\phi \equiv \Phi - {1\over 2} {\rm ln}\,{\rm det}\bar{e}^a_m$ is
the four-dimensional dilaton and $\bar{\sigma} \equiv {\rm ln}\,{\rm det}
\bar{e}^a_m$ parameterizes the volume of a six-torus,
$\bar{g}_{mn} \equiv \eta_{ab}\bar{e}^a_m \bar{e}^b_n$, and
$\bar{G}^m_{\mu\nu} \equiv \partial_{\mu} \bar{B}^m_{\mu} -
\partial_{\nu} \bar{B}^m_{\mu}$.  Here, the field strengths
$\bar{F}_{\mu\nu\,m}$, $\bar{G}_{\mu\nu}$ and $\bar{F}_{\mu\nu\,mn}$
are defined in terms of the Abelian gauge fields decomposed from
the ten-dimensional two-form $A_{\breve{\mu}\breve{\nu}}$, the one-form
$B_{\breve{\mu}}$ and the three-form $A_{\breve{\mu}\breve{\nu}
\breve{\rho}}$ fields, respectively.  The following Einstein-frame
action can be obtained by the Weyl rescaling $g_{\mu\nu} \to g^E_{\mu\nu}
= e^{-2\phi}g_{\mu\nu}$:
\begin{eqnarray}
{\cal L}_{II} = &-&{1\over 4}e^E[{\cal R}^E - 2\partial_{\mu}\phi
\partial^{\mu} \phi + {1\over 4}\partial_{\mu}\bar{g}_{mn}
\partial^{\mu}\bar{g}^{mn} - {1\over 4}e^{-2\phi}\bar{g}_{mn}
\bar{G}^m_{\mu\nu}\bar{G}^{n\,\mu\nu}-e^{-2\phi}\bar{g}^{mn}
{\bar F}_{\mu\nu\, m}{\bar F}^{\mu\nu}_{\ \ n}
\nonumber \\
&+& {1\over 4}e^{\bar \sigma}\bar{G}_{\mu\nu}\bar{G}^{\mu\nu}
+ {1\over 2}e^{\bar \sigma}\bar{g}^{mn}\bar{g}^{pq}
\bar{F}_{\mu\nu\, mp}\bar{F}^{\mu\nu}_{\ \ nq}].
\label{eintype2}
\end{eqnarray}

One has to notice here that when we relate the fields of
eleven-dimensional supergravity to those of type-IIA supergravity
on a torus, we kept only scalar fields that are associated only
with the Elfbein $E^{(11)\,A}_M$ (thereby, breaking $E_7$ to $SL(7)$),
and turned off $g^{(11)}_{m7}$ or, equivalently, $B_m$, (thereby,
breaking $SL(7)$ to $SL(6)$).  Therefore, the symmetry
transformations of the four-dimensional Lagrangian (\ref{eintype2})
do not mix RR $U(1)$ and NS-NS $U(1)$ charges.
Since the scalar fields associated with $g^{(11)}_{m7}$ are turned off,
the $SO(7)$ symmetry of the eleven-dimensional supergravity on $T^7$ is
broken down to the $SO(6)$ symmetry.  In the RR sector, the $U(1)$ gauge
fields $\bar{B}_{\mu}$ and $\bar{A}_{\mu\,mn}$ transform under $SO(6)$
as a singlet and $\bf 15$ antisymmetric representation,
respectively.  In the NS-NS sector, $\bar{A}_{\mu\,mn}$ and
$\bar{B}^m_{\mu}$ belong to $\bf 15$ antisymmetric and $\bf 6$
vetor representations of $SO(6)$, respectively.

The eleven-dimensional supergravity on $T^7$ and the type-IIA
supergravity on $T^6$ are described by the same $N=8$ supergravity
in four dimensions.  In fact, the former is the strong coupling
limit \cite{WITTEN} of the later.  The bosonic fields in (\ref{eintype2})
are related to those in (\ref{action4d}) by:
\begin{eqnarray}
\phi &=& -{3\over 7}{\varphi} + {1 \over 4}{\rm ln}\, \rho_{77},\ \ \
\bar{\sigma} = {9\over 7}\varphi + {\rm ln} \, \rho_{77} ,\ \ \
\bar{\rho}_{mn} = (\rho_{77})^{1\over 6}\rho_{mn},
\nonumber \\
\bar{B}^m_{\mu} &=& B^m_{\mu}, \ \ \ \  \bar{B}_{\mu} = B^7_{\mu} , \ \ \
\bar{A}_{\mu m} = A_{\mu m 7}, \ \ \ \  \bar{A}_{\mu\,mn} = A_{\mu\,mn},
\label{rel}
\end{eqnarray}
where $m,n = 1,...,6$ and $\bar{\rho}_{mn}$ is the unimodular
part of the internal metric $\bar{g}_{mn}$ ($\bar{g}_{mn} =
-e^{\bar{\sigma}/3}\bar{\rho}_{mn}$).

By using the relations (\ref{rel}), one can obtain the dyonic BH
solutions carrying charges of each $U(1)$ fields in the RR and/or
the NS-NS sector(s) of the type-IIA superstring.  In the following, we
shall classify the type-IIA superstring BH solutions according to the
type of eleven-dimensional field, {\it i.e.}, $E^{(11)\,A}_M$ and
$A^{(11)}_{MNP}$, from which the four-dimensional $U(1)$ gauge fields
are originated, and summarize the dependence of the ADM masses of
these solutions on the asymptotic values of four-dimensional scalars.
Note, the string frame ADM mass is related to the Einstein frame ADM
mass as $M_s = e^{-\phi_{\infty}}M_E$.

The first class of solutions corresponds to black holes carrying
charges of $U(1)$ gauge fields associated with the off-diagonal
components of the eleven-dimensional space-time metric:
\begin{itemize}
\item
{\bf Type-KNR solutions\ } Magnetic charge P associated with
$\bar{B}_{\mu}$, {\it i.e.}, the one-form $U(1)$ field in the RR sector,
and electric charge $Q_m$ associated with $\bar{B}^m_{\mu}$, {\it i.e.},
one of six Kaluza-Klein gauge fields in the NS-NS sector:\\
\begin{eqnarray}
e^{(\phi-\phi_{\infty})} &=& \left ({{r-{\bf P}_\infty-{\bf
Q}_{m\,\infty}} \over {r-{\bf P}_\infty}} \right )^{1\over 4}, \ \
e^{2(\bar{\sigma}-\bar{\sigma}_{\infty})} =  {{(r -{\bf P}_\infty-{\bf
Q}_{m\,\infty})^2} \over {(r -{\bf P}_\infty)^ {-1}(r-{\bf
Q}_{m\,\infty})^3}},
\nonumber \\
\bar{\rho}_{mm}/\bar{\rho}_{mm\,\infty}&=&\left ( {{r-{\bf P}_\infty}\over
{r -{\bf P}_\infty-{\bf Q}_{m\,\infty}}} \right )^{5\over 6}, \ \
\bar{\rho}_{kk}/\bar{\rho}_{kk\,\infty} = \left ({{r-{\bf P}_\infty -{\bf
Q}_{m\,\infty}} \over {r-{\bf P}_\infty}} \right )^ {1\over 6}\ \ (k \neq m),
\nonumber \\
M_E &=& |{\bf P}_\infty|+|{\bf Q}_{m\,\infty}|=
e^{\bar{\sigma}_\infty/2}|P|+e^{-\phi_\infty}\bar{g}^{1\over 2}_{mm\,\infty}
|Q_m|.
\label{Kaluza-Kleinnr}
\end{eqnarray}
The $SO(6)/SO(5)$ rotations on this solution induce ${{6\cdot 5}\over 2}
-{{5\cdot 4}\over 2} = 5$ new magnetic charge degrees of freedom in the
gauge fields $\bar{B}_m$.  For the case where electric charge $Q$
and magnetic charge $P_m$ are associated with $B_{\mu}$ and
$\bar{B}^m_{\mu}$, respectively, one can obtain the corresponding
solutions by imposing the electric-magnetic duality transformations.
\item
{\bf Type-KNN solutions\ } Magnetic charge $P_m$ associated with
$\bar{B}^m_{\mu}$ and electric charge $Q_n$ associated with
$\bar{B}^n_{\mu}$, {\it i.e.}, both charges correspond to Kaluza-Klein
$U(1)$ fields of the NS-NS sector:\\
\begin{eqnarray}
e^{(\phi-\phi_{\infty})} &=&
\left ( {{r-{\bf Q}_{n\,\infty}} \over  {r-{\bf P}_{m\,\infty}}}
 \right)^{1\over 4}, \ \  e^{2(\bar{\sigma}-\bar{\sigma}_{\infty})} =
{{r-{\bf P}_{m\,\infty}} \over {r-{\bf Q}_{n\,\infty}}}, \ \
\bar{\rho}_{mm}/\bar{\rho}_{mm\,\infty} =  {{r-{\bf P}_{m\,\infty}-{\bf
Q}_{n\,\infty}} \over  {(r-{\bf P}_{m\,\infty})^{1\over 6}(r-{\bf
Q}_{n\,\infty})^{5\over 6}}},
\nonumber \\
\bar{\rho}_{nn}/\bar{\rho}_{nn\,\infty} &=& {{(r-{\bf P}_{m\,\infty})^
{5\over 6}(r-{\bf Q}_{n\,\infty})^{1\over 6}}\over {r-{\bf P}_{m\,\infty}
-{\bf Q}_{n\,\infty}}}, \ \
\bar{\rho}_{\ell\ell}/\bar{\rho}_{\ell\ell\,\infty} =  \left ( {{r-{\bf
Q}_{n\,\infty}} \over {r-{\bf P}_{m\,\infty}}}  \right )^{1\over 6}\ \ \
(\ell \neq m,n),
\nonumber \\
M_E &=& |{\bf P}_{m\,\infty}|+|{\bf Q}_{n\,\infty}|=
e^{-\phi_\infty} \bar{g}^{1\over 2}_{mm\,\infty}|P_m| +
e^{-\phi_\infty}\bar{g}^{1\over 2}_{nn\,\infty}|Q_n|,
\label{Kaluza-Kleinnn}
\end{eqnarray}
Upon imposing the $SO(6)/SO(4)$ rotations, one has the most general
supersymmetric eleven-dimensional Kaluza-Klein BH's with the constraint
$\sum P_i Q_i = 0$.
\end{itemize}

Secondly, we have the following classes of dyonic solutions that
correspond to $U(1)$ gauge fields associated with the eleven-dimensional
three-form field $A^{(11)}_{MNP}$:
\begin{itemize}
\item
{\bf Type-HNR solutions\ } Magnetic charge $P_m$ associated with
$\bar A_{{\mu}\,m}$, {\it i.e.}, one of six 2-form $U(1)$ fields
in the NS-NS sector, and the electric charge $Q_{mn}$ associated with
$\bar{A}_{\mu\,mn}$, {\it i.e.}, one of fifteen three-form $U(1)$ fields
in the RR sector:\\
\begin{eqnarray}
e^{(\phi-\phi_{\infty})} &=& \left ( {{r-2{\bf Q}_{mn\,\infty}} \over
{r-2{\bf P}_{m\,\infty}-2{\bf Q}_{mn\,\infty}}} \right )^{1\over 4}, \ \
e^{2(\bar{\sigma}-\bar{\sigma}_{\infty})} =  {{(r-2{\bf P}_{m\,\infty})
(r-2{\bf Q}_{mn\,\infty})} \over  {(r - 2{\bf P}_{m\,\infty}-2{\bf Q}_
{mn\,\infty})^2}},
\nonumber \\
\bar{\rho}_{mm}/\bar{\rho}_{mm\,\infty} &=& {{(r-2{\bf P}_{m\,\infty})^
{-{2\over 3}}(r-2{\bf Q}_{mn\,\infty})^{5\over 6}} \over {(r-2{\bf P}_
{m\,\infty}-2{\bf Q}_{mn\,\infty})^{1\over 6}}}, \ \
\bar{\rho}_{nn}/\bar{\rho}_{nn\,\infty} = {{(r-2{\bf P}_{m\,\infty}-
2{\bf Q}_{mn\,\infty})^{5\over 6}} \over  {(r-2{\bf P}_{m\,\infty})^
{2\over 3}(r-2{\bf Q}_ {mn\,\infty})^{1\over 6}}},
\nonumber \\
\bar{\rho}_{\ell\ell}/\bar{\rho}_{\ell\ell\,\infty} &=& {{(r-2{\bf P}_
{m\,\infty})^{1\over 3}(r-2{\bf Q}_{mn\,\infty})^{-{1\over 6}}}
\over {(r-2{\bf P}_{m\,\infty} - 2{\bf Q}_{mn\,\infty})^{1\over 6}}}\ \
(\ell \neq m,n),
\nonumber \\
M_E &=& 2|{\bf P}_{m\,\infty}|+2|{\bf Q}_{mn\,\infty}|=
2e^{-\phi_\infty} \bar{g}^{-{1\over 2}}_{mm\,\infty}|P_m| +
2e^{\bar{\sigma}_\infty/2}\bar{g}^{-{1\over 2}}_{mm\,\infty}
\bar{g}^{-{1\over 2}}_{nn\,\infty}|Q_{mn}|.
\label{threenr}
\end{eqnarray}
The $SO(6)/SO(4)$ rotations induce ${{6\cdot 5}\over 2} - {{4\cdot 3}
\over 2} = 9$ new charge degrees of freedom.  For the case of
electric charge $\bar Q_m$ coming from $\bar A_{\mu\,m}$ and
magnetic charge $P_{mn}$ coming from $A_{\mu\,mn}$, the corresponding
solutions can be obtained by imposing the electric-magnetic duality
transformations.
\item
{\bf Type-HRR solutions\ }  Magnetic charge $P_{mn}$ coming from
$A_{\mu\,mn}$ and electric charge $Q_{mp}$ coming from $A_{\mu\,mp}$,
both of which are the charges of 3-form $U(1)$ fields in the R-R
sector:\\
\begin{eqnarray}
e^{(\phi-\phi_{\infty})}&=&1, \ \  e^{2(\bar{\sigma}-\bar{\sigma}_{\infty})}
=  {{r - 2{\bf P}_{mn\,\infty}} \over {r - 2{\bf Q}_{mp\,\infty}}},
\nonumber \\  \bar{\rho}_{mm}/\bar{\rho}_{mm\,\infty} &=&
\left ({{r-2{\bf P}_{mn\,\infty}} \over {r-2{\bf Q}_{mp\,\infty}}}\right )^
{-{2\over 3}}, \ \  \bar{\rho}_{nn}/\bar{\rho}_{nn\,\infty} =
{{(r-2{\bf P}_{mn\,\infty})^{1\over 3}(r-2{\bf Q}_{mp\,\infty})^
{2\over 3}}  \over {r-2{\bf P}_{mn\,\infty}-2{\bf Q}_{mp\,\infty}}},
\nonumber \\
\bar{\rho}_{pp}/\bar{\rho}_{pp\,\infty} &=&  {{r - 2{\bf P}_{mn\,\infty}-
2{\bf Q}_{mp\,\infty}} \over  {(r-2{\bf P}_{mn\,\infty})^{2\over 3}
(r-2{\bf Q}_ {mp\,\infty})^{1\over 3}}}, \ \
\bar{\rho}_{\ell\ell} = \left ({{r-2{\bf P}_{mn\,\infty}} \over
{r-2{\bf Q}_{mp\,\infty}}} \right )^{1\over 3}\ \ \ (\ell \neq m,n,p),
\nonumber \\
M_E &=& 2|{\bf P}_{mn\,\infty}|+2|{\bf Q}_{mp\,\infty}|=
2e^{\bar{\sigma}_\infty/2}\bar{g}^{-{1\over 2}}_{mm\,\infty}
\bar{g}^{-{1\over 2}}_{nn\,\infty} |P_{mn}| +
2e^{\bar{\sigma}_\infty /2}\bar{g}^{-{1\over 2}}_{mm\,\infty}
\bar{g}^{-{1\over 2}}_{pp\,\infty} |Q_{mp}|,
\label{threenn}
\end{eqnarray}
The $SO(6)/SO(2)$ transformations on this solution introduce
${{6\cdot 5}\over 2}-{{2\cdot 1}\over 2} = 14$ charge degrees of
freedom in the gauge fields $A_{\mu\,ij}$.
\item
{\bf Type-HNN solutions\ } Magnetic charge $P_m$ associated with
$\bar{A}_{\mu\,m}$ and electric charge $Q_n$ associated with
$\bar{A}_{\mu\,n}$, {\it i.e.}, both charges arise from 2-form
$U(1)$ fields in the NS-NS sector
\footnote{Note, that for a special case with either $\bar{P}_m=0$ or
$\bar{Q}_n=0$, the result reduces to a solution first found in Ref.
\cite{BANKS}, and it is related to $H$-monopoles of the heterotic
string \cite {KHURI,HARVEY}.}: \\
\begin{eqnarray}
e^{(\phi-\phi_\infty)} &=&
\left ({{r-2{\bf Q}_{n\,\infty}}\over{r-2{\bf P}_{m\,\infty}}}\right )^
{1\over 4}, \ \
e^{2(\bar{\sigma}-\bar{\sigma}_\infty)} = {{r-2{\bf Q}_{n\,\infty}}
\over {r-2{\bf P}_{m\,\infty}}},
\nonumber \\
\bar{\rho}_{mm}/\bar{\rho}_{mm\,\infty} &=& {{(r-2{\bf P}_{m\,\infty})^
{1\over 6}(r-2{\bf Q}_{n\,\infty})^{5\over 6}} \over{r-2{\bf P}_{m\,\infty}
- 2{\bf Q}_{n\,\infty}}}, \ \
\bar{\rho}_{nn}/\bar{\rho}_{nn\,\infty} = {{r-2{\bf P}_{m\,\infty}
-2{\bf Q}_{n\,\infty}} \over {(r-2{\bf P}_{m\,\infty})^{5\over 6}
(r-2{\bf Q}_{n\,\infty})^{1\over 6}}},
\nonumber \\
\bar{\rho}_{\ell\ell}/\bar{\rho}_{\ell\ell\,\infty} &=&
\left ( {{r-2{\bf P}_{m\,\infty}} \over {r-2{\bf Q}_{n\,\infty}}}
\right )^{1\over 6} \ \ \ (\ell \neq m,n),
\nonumber \\
M_E &=& 2|{\bf P}_{m\,\infty}|+2|{\bf Q}_{n\,\infty}|=
2e^{-\phi_\infty}\bar{g}^{-{1\over 2}}_{mm\,\infty}|P_m| +
2e^{-\phi_\infty} \bar{g}^{-{1\over 2}}_{nn\,\infty}|Q_n|.
\label{threerr}
\end{eqnarray}
Upon imposing the $SO(6)/SO(4)$ rotations, one obtains the most general
supersymmetric four-dimensional 2-form BH solutions with the constraint
$\sum Q_i P_i=0$.
\end{itemize}
In the above, the expressions for the four-dimensional metric components
are given by $\lambda(r)$ and $R(r)$ of either (\ref{Kaluza-Kleinsol}) or
(\ref{antisol}) with the corresponding charges in the NS-NS or the RR
sector replaced.

In the expressions for the Einstein frame ADM mass $M_E$ [string frame
ADM mass $M_s=e^{-\phi_{\infty}}M_E$], the screened charges from the RR
sector do not scale [scale as $e^{-\phi_{\infty}}$] with respect to the
asymptotic string coupling $e^{\phi_{\infty}}$, while the screened
charges from the NS-NS sector scale as $e^{-\phi_{\infty}}$ [scale as
$e^{-2\phi_{\infty}}$], in agreement with a general analysis in Ref.
\cite{WITTEN}.  Note also the following scaling dependence of screened
charges on the asymptotic volume of the six-torus, parameterized by
$e^{\bar{\sigma}_{\infty}}$: the screened charges, associated with the
$U(1)$ fields in the RR and the NS-NS sectors, scale as
$e^{\bar{\sigma}_{\infty}/2}$ and do not scale, respectively.

These classes of solutions do not become massless at any point of
moduli space or the choice of the charge lattice except at the
boundary of the moduli space, {\it i.e.}, $g_{mm\,\infty} = \pm\infty$,
or in the strong coupling limit, {\it i.e.}, $e^{\phi_{\infty}} = \infty$,
of the string theory.

The massless solutions arise for $N=4$ string vacua \cite{CYHETS}
when BPS saturated states carry charges of both, the Kaluza-Klein $U(1)$
field and the two-form $U(1)$ field in the NS-NS sector, which is the
common sector of the type-IIA and the heterotic string theory.
In Ref. \cite{HTII}, it was argued that in general $N=8$ theory in four
dimensions, {\it i.e.}, the eleven-dimensional supergravity on a
seven-torus or the type-IIA superstring on a six-torus, does not give
rise to massless BPS saturated states for non-trivial charge configurations.
However, the latter observation is not in contradiction with the fact
that the BPS saturated states carrying charges of the NS-NS sector, which
is a part of the type-IIA as well as the heterotic superstring, can
become massless for particular charge configurations at special points of
moduli space.  Namely, in order to turn off the $U(1)$ charges of the
RR sector of the type-IIA superstring, one has to discard one chirality of
the spinor in order to ensure the consistency of the supersymmetry
transformations for the corresponding bosonic fields, thus ensuring that in
this case the theory possesses only $N=4$ supersymmetry in four
dimensions.

\section{Conclusion}

We have discussed two classes of dyonic black hole solutions in
the eleven-dimensional supergravity on a seven-torus, {\it i.e.},
one associated with the Kaluza-Klein $U(1)$ fields coming from the
off-diagonal components of the Elfbein $E^{(11)\,A}_M$ and the other
one, associated with the three-form $U(1)$ fields arising from
the three-form field $A^{(11)}_{MNP}$.
These two classes of solutions have the same four-dimensional
space-time property, since all of the $U(1)$ gauge fields and
their duals are related through the global $E_7$ symmetry, which
leaves the Einstein frame four-dimensional metric intact
\footnote{Note that the $E_7$ symmetry puts all the scalars on
the same footing.  So, the dilaton mixes with other scalars under
the $E_7$ transformation.}.
It is a subject of future investigation to explore the full $E_7$
symmetry of the BPS saturated states in the four-dimensional
$N=8$ supergravity theory and its relationship to the full BPS
soliton spectrum of the type-IIA superstring on a six-torus.

Since the eleven-dimensional supergravity on a seven-torus and
the type-IIA supergravity on a six-torus are described by the
four-dimensional $N=8$ supergravity, which is unique, the
effective Lagrangian of these two theories are related through
the field redefinition.  In fact, the former is the strong coupling
limit of the latter.  In this paper, we obtained the field redefinitions
for the special case when the original $E_7$ symmetry of the Lagrangian
is broken down to $SL(6)$, {\it i.e.}, all the other scalar fields except
those associated with $e^{(10)\,a}_m$ and $\Phi$ are turned off.
By using the field redefinitions we have obtained dyonic black hole
solutions carrying charges of the RR and/or the NS-NS sector(s) of
the type-IIA superstring from the dyonic solutions carrying either
KK or the 3-form $U(1)$ charges of the eleven-dimensional
supergravity on a seven-torus.
The ADM masses of these configurations scale with the asymptotic
string couplings in accordance with the general analysis in
Ref. \cite{WITTEN}.  In addition, the scaling of the ADM mass
with the asymptotic value of the volume of the six-torus is
discussed.

The classes of solutions we have discussed in this work are a
special case of more general ``generating'' solution, which
would induce all the BPS saturated solutions after imposing
a subset of $E_7$ group.  This ``generating'' solution will
carry electric charges and magnetic charges, each of which coming
from Kaluza-Klein $U(1)$ and three-form $U(1)$ fields (of the
eleven-dimensional supergravity on a seven-torus), respectively.
More general classes of solutions, including non-supersymmetric
as well as BPS saturated solutions, can be obtained by introducing
another parameter (into the generating solutions of the most general
BPS states) of the $E_8$ group, which is the symmetry of the
three-dimensional effective action for stationary solutions of the
eleven-dimensional supergravity.

By explicitly solving the Killing spinor equations, we have shown that
classes of dyonic extreme black hole solutions carrying the $U(1)$
charges of the NS-NS and/or the RR sectors of the type-IIA superstring
preserve part of supersymmetries in the four-dimensional vacua of the
effective supergravity theory.  Purely electrically (or purely
magnetically) charged solutions preserve $1\over 2$ of the
original supersymmetries, thereby forming short multipletes, and
dyonic solutions preserve $1\over 4$ of the original supersymmetries,
thereby forming intermediate multipletes.  The ultimate goal is to
obtain the most general BPS solutions in the $N=8$ vacua of the
superstring theories.

\acknowledgments
The work is supported by the Institute for Advanced Study funds and
J. Seward Johnson foundation,  U.S. Department of Energy Grant No.
DOE-EY-76-02-3071, the National Science Foundation Career Advancement
Award PHY95-12732 and the NATO collaborative research grant CGR 940870.


\vskip2.mm



\begin{references}


\bibitem{STRWK}{J.H. Schwarz, CALT-68-1815 preprint (1992), hep-th \#
9209125.}

\bibitem{HETEM}{A. Shapere, S. Trivedi and F. Wilczek, Mod. Phys. Lett.
{\bf A6} (1991) 2677; A. Sen, Nucl. Phys. {\bf B404} (1993) 109.}

\bibitem{UDUAL}{C.M. Hull and P.K.Townsend, Nucl. Phys. {\bf B438}, (1995)
109.}

\bibitem{DSZ}{P. Dirac, Proc. R. Soc. {\bf A133} (1931) 60; J. Schwinger,
Phys. Rev. {\bf 144} (1966) 1087; {\bf 173} (1968) 1536; D. Zwanzinger,
Phys. Rev. {\bf 176} (1968) 1480, 1489.}

\bibitem{TDUAL}{A. Giveon, M. Porrati and E. Rabinovici, Phys. Rep.
{\bf 244} (1994) 77.}

\bibitem{SENS}{A. Sen, Int. J. Mod. Phys. {\bf A9} (1994) 3707.}

\bibitem{NARAIN}{K. Narain, Phys. Lett. {\bf B169} (1986) 41;
K. Narain, H. Sarmadi and E. Witten, Nucl. Phys. {\bf B279}
(1987) 369.}

\bibitem{WITTEN}{E. Witten, Nucl. Phys. {\bf B443} (1995) 85.}

\bibitem{BPS}{E.B. Bogomol'nyi, Sov. J. Nucl. Phys, {\bf 24} (1976) 449;
M.K. Prasad and C.M. Sommerfield, Phys. Rev Lett. {\bf 35} (1975) 760.}

\bibitem{HULLCONF}{C.M. Hull,  Talk given at STRINGS'95: Future
Perspectives in String Theory, Los Angeles, CA, 13-18 Mar 1995.}

\bibitem{STROM}{A. Strominger, Nucl. Phys. {\bf B451} (1995) 96.}

\bibitem{HTII}{C.M. Hull and P.K. Townsend, QMW-95-10 preprint, hep-th\#
9505073.}

\bibitem{CYHETS}{M. Cveti\v c and D. Youm, UPR-674-T preprint,
hep-th \# 9507160, to be published in  Phys. Lett. {\bf B}.}

\bibitem{CHNUL}{A.A. Tseytlin, IMPERIAL-TP-94-95-28, hep-th\# 9505052,
and references therein.}

\bibitem{CYCONF}{M. Cveti\v c and D. Youm, UPR-675-T (1995)
hep-th\# 9508058, Talk given at STRINGS'95: Future Perspectives in
String Theory, Los Angeles, CA, 13-18 Mar 1995.}

\bibitem{ALLHET}{M. Cveti\v c and D. Youm, UPR-672-T, hep-th\# 9507090.}

\bibitem{TOWN2}{P. K. Townsend, DAMTP-R-95-2, hep-th\# 9501068, Phys. Lett.
{\bf B350} (1995) 184.}

\bibitem{ELEV}{M. Cveti\v c and D. Youm, UPR-658-T, hep-th\# 9505045,
to be published in  Phys. Rev. {\bf D}.}

\bibitem{HUQ}{M. Huq and M.A. Namazie, Class. Quant. Grav. {\bf 2}
(1985) 293.}

\bibitem{ELE}{E. Cremmer and B. Julia, Phys. Lett. {\bf 76B} (1978) 409.}

\bibitem{CRE}{E. Cremmer and B. Julia, Nucl. Phys. {\bf B159} (1979) 141.}

\bibitem{SUPER}{M. Cveti\v c and D. Youm, hep-th \# 9409119, Nucl. Phys.
{\bf  B438} (1995) 182, and addendum UPR-650-T, hep-th \# 9503081,
Nucl. Phys. {\bf B449} (1995) 146.}

\bibitem{STAT}{M. Cveti\v c and D. Youm, hep-th\# 9502099, Phys. Rev. {\bf
D52} (1995) 2144.}

\bibitem{BANKS}{T. Banks, M. Dine, H. Dykstra and W. Fischler, Nucl.
Phys. {\bf B212} (1988) 45.}

\bibitem{KHURI}{R. Khuri, Phys. Lett.  {\bf 259B} (1991) 261;
Phys. Lett. {\bf 294B} (1992) 325; Nucl. Phys. {\bf B387} (1992) 315.}

\bibitem{HARVEY}{J.A. Harvey and J. Liu, Phys. Lett. {\bf B268} (1991) 40.}


\end{references}
\end{document}